# Design of Mechanically Compliant Membrane Reflectors for Giant Optical Phase Nonlinearity

**LIOR MICHAELI[1]* AND HARRY A. ATWATER[2]**

[1]*School of Electrical and Computer Engineering, Faculty of Engineering, Tel Aviv University, 6997801 Tel Aviv, Israel*
[2]*Department of Applied Physics and Materials Science, California Institute of Technology, Pasadena, California 91125, USA*
**Corresponding author: liormic1@tauex.tau.ac.il*

**Abstract**

Optical nonlinearities arise when the response of a photonic system depends on light intensity. Radiation pressure can create such a nonlinearity by moving a mechanically compliant reflector, thereby shifting the phase of the reflected light. Based on measured properties, we predict a giant phase responsivity of $263\,\mathrm{rad\,W^{-1}}$ and a device-equivalent $n_{2,\mathrm{eff}} = 5.4 \times 10^{-7}\,\mathrm{m^2\,W^{-1}}$ for a silicon nitride membrane trampoline with serpentine springs. We introduce strength-range-aperture metrics for reflective phase elements. Among reported mechanical resonators compared under a common direct-reflection protocol, this trampoline is an outlier, combining high compliance, practical optical accessibility, and 99.85% retention of its small-signal phase responsivity through a full $2\pi$ reflected-phase shift. Practical implementation requires optical and thermal co-design. Our framework indicates design priorities for future mechano-optical phase elements.

## 1. Introduction

Nonlinear optical phenomena are central to modern photonics, enabling key functions in communications, signal processing, sensing, and optical computing [1–5]. Established nonlinear mechanisms differ in strength, speed, loss, and spectral bandwidth. For example, electronic Kerr effects are fast and reversible but typically weak [1,6]. Resonant nanophotonic structures enhance nonlinear interactions but introduce spectral selectivity [7,8]. Mechanisms based on molecular reorientation, thermo-optic response, and phase transitions can produce large phase shifts, but may variously involve high optical loss, slow dynamics, or hysteresis [6,9,10]. Frequency-conversion processes such as harmonic generation and parametric mixing provide powerful functionality, but generate optical fields at new frequencies rather than acting solely on the original carrier wave [1]. These trade-offs become especially important in cascaded optical systems [5,11–13], where optical losses accumulate across successive stages, while frequency-converting nonlinearities do not preserve a common carrier frequency throughout such a cascade. For cascaded operation, nonlinear phase control should therefore ideally be strong, low-loss, reversible, frequency-preserving, and broadband across optical wavelengths. A reliable quantitative model of its response is also needed to predict and design the behavior of the full cascade.

Optical forces provide a distinct route to nonlinearity by moving or deforming a photonic structure and thereby modifying the field, a principle central to cavity optomechanics [14]. This approach has enabled large reconfigurable responses in nanomechanical metamaterials and optomechanical phase shifters across free-space and integrated platforms [15–18]. Among these approaches, a movable reflector provides a particularly direct route to phase modulation. Radiation pressure displaces the reflecting boundary, and the resulting optical path change is encoded onto the reflected field. In the quasi-static limit, the interaction requires neither wavelength conversion nor a narrow optical resonance. Its usable wavelength range is governed mainly by the reflector's constitutive optical properties, while its temporal response is governed by the mechanical dynamics.

To assess mechanical resonators as optically driven phase elements, we introduce a framework that separates their mechanical and optical properties. Mechanically, compliance determines the displacement produced by a small optical force. Yet useful phase control also requires this response to be retained over the displacement needed for a large phase shift. Optically, the phase-coupled area measures the effective portion of the illuminated region that both drives the mechanical motion and appears as a common phase shift in the reflected light. Taken together, the small-signal compliance, retained-response range, and phase-coupled area define the intrinsic phase-modulation potential used in the comparison below.

The silicon nitride trampoline in Fig. 1(a) provides a direct test case for this framework. Its large reflective pad is suspended by compliant serpentine springs, combining a practical optical aperture with exceptionally low stiffness. We introduced this architecture for calibrated radiation-pressure measurements on membrane lightsails [19]. The related device shown here was subsequently characterized in the large-amplitude Duffing regime [20], providing the measured linear stiffness and cubic stiffness coefficient used below.

**a** Compliant membrane

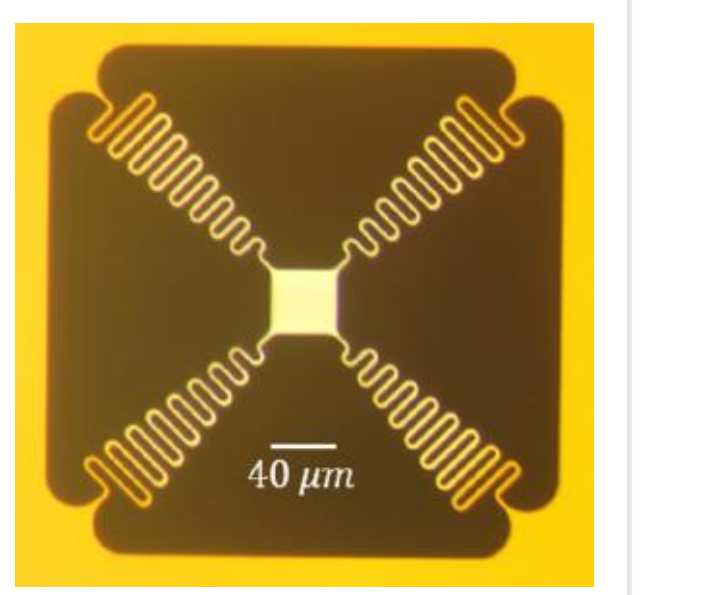

**b** Radiation-pressure-induced phase nonlinearity

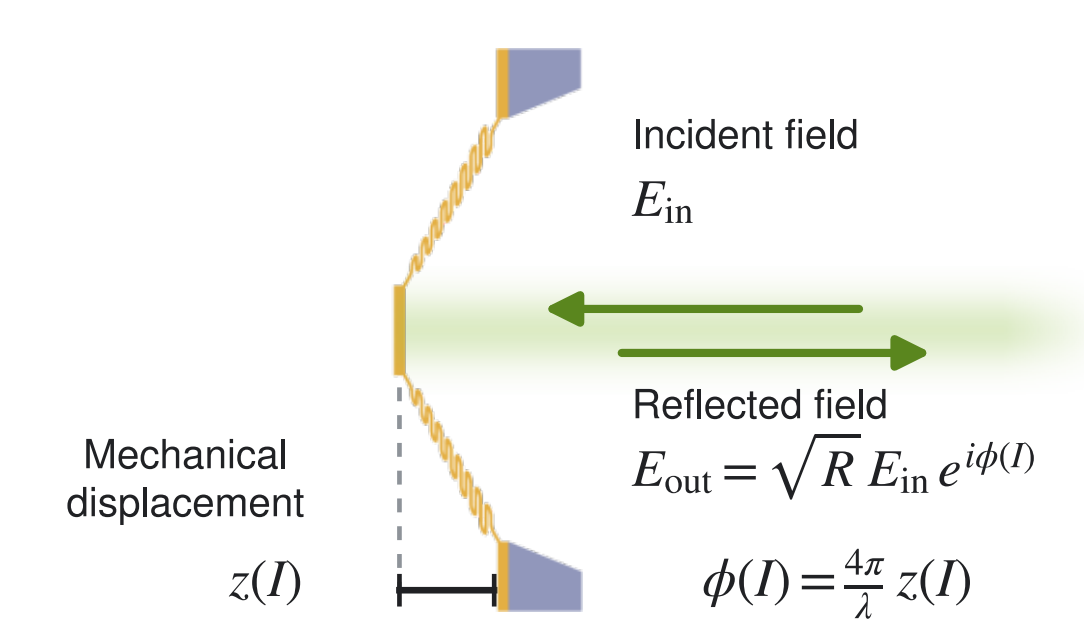

Fig. 1. Compliant membrane for radiation-pressure-induced optical phase nonlinearity. (a) Optical micrograph of the 50-nm-thick silicon nitride trampoline resonator. The serpentine suspension provides high compliance and a large retained-compliance range, while the central pad provides nearly uniform normal motion across a large reflective interface. Scale bar, 40 µm. Micrograph also shown in Ref. [20]. (b) In the quasi-static, uniform-motion limit, radiation pressure displaces the membrane by $z(I)$, producing a reflected phase shift $\varphi(I)$=$4\pi z(I)/\lambda$.

We next use this framework to place the trampoline within the context of a broader performance landscape for direct-reflection phase modulation. The quantitative comparison includes fundamental or static-dominant nano- and micromechanical systems for which the required mechanical response and spatial information can be reconstructed from data in the cited studies under a common illumination and readout geometry. This common reconstruction protocol places all plotted systems on the same quantitative basis.

Such a comparison reveals that strong mechanical performance does not necessarily translate into strong optical phase modulation. Several nanoscale resonators reported previously can match or exceed our membrane trampoline in compliance or retained range. Yet their small phase-coupled areas limit how effectively radiation pressure can drive that mechanical response and convert it into a reflected phase shift. With wavelength and the optical momentum-transfer coefficient $\xi$ held common, the projected small-signal slope of reflected phase versus incident intensity for the membrane trampoline is about 2,000 times that of the next-highest system meeting the reconstruction criteria for both axes. Combining optical data from Ref. [19] with mechanical data from Ref. [20], we predict a device-equivalent nonlinear index of $n_{2,\mathrm{eff}} = 5.4 \times 10^{-7}\ \mathrm{m}^2\,\mathrm{W}^{-1}$ and 99.85% responsivity retention through a $2\pi$ phase shift. The retained range defines the available phase potential, while optical momentum transfer and thermal management determine how much of this range can be accessed in operation. These results indicate that membrane trampolines represent a new benchmark and can drive the design criteria for future mechano-optical phase elements.

## 2. From radiation pressure to usable optical phase nonlinearity

Consider a normally incident drive beam of power $P_d$ and wavelength $\lambda$ illuminating a movable reflector. Let $z$ denote its normal displacement from equilibrium at $P_d = 0$. For spatially uniform normal motion, $z$ changes the reflected optical path by $2z$, while reflection and absorption transfer momentum to the structure. The reflected phase shift and total optical force are

$$\phi = \frac{4\pi z}{\lambda}, \qquad F_{\text{opt}} = \frac{\xi P_d}{c}, \qquad \xi \equiv 2R + A_{\text{abs}}. \tag{1}$$

Here $R$ and $A_{\text{abs}}$ are the reflectance and absorptance, $c$ is the speed of light in vacuum, and $\xi$ is the optical-momentum-transfer coefficient. Equation (1) summarizes the uniform-motion limit shown in Fig. 1(b) and the sequence from incident power to optical force, mechanical displacement, and reflected phase.

A mechanical mode may vary across the reflecting surface. Let $S$ denote the surface included in the calculation, $\mathbf{r}$ a position on that surface, and $dA$ an infinitesimal surface element. We write the local normal displacement of the membrane as $u_z(\mathbf{r}) = z\psi(\mathbf{r})$. The dimensionless mode shape $\psi(\mathbf{r})$ describes the spatial motion, while $z$ sets its amplitude. We choose $\max_{\mathbf{r}\in S}|\psi(\mathbf{r})| = 1$, so that $|z|$ is the largest normal displacement on the surface.

The optical field has two roles. It applies radiation pressure to the mechanical mode, and it carries the resulting motion into the measured reflected phase. The normalized drive profile is $w_d(\mathbf{r}) = I_d(\mathbf{r})/P_d$, where $I_d(\mathbf{r})$ is the local incident intensity and $P_d = \int_S I_d(\mathbf{r})\, dA$ is the total drive power. The normalized readout profile $w_r(\mathbf{r})$ weights the phase across the reflected surface in the collected optical mode. Both profiles integrate to unity, $\int_S w_j(\mathbf{r})\, dA = 1$ for $j \in \{d, r\}$, and are zero outside their corresponding illuminated or collected regions.

Projecting the spatial optical force onto the chosen mechanical coordinate, and linearizing the collected reflected phase with respect to displacement, gives

$$\begin{aligned} \Gamma_j &\equiv \int_S w_j(\mathbf{r})\psi(\mathbf{r})\, dA, \qquad j \in \{d, r\}, \\ F_{\text{opt}} &= \frac{\xi P_d}{c}\Gamma_d, \qquad \frac{d\phi}{dz} = \frac{4\pi}{\lambda}\Gamma_r. \end{aligned} \tag{2}$$

The two overlaps describe the optical coupling into and out of the selected mechanical coordinate. The drive overlap $\Gamma_d$ quantifies how effectively the spatial radiation-pressure force drives that coordinate. The readout overlap $\Gamma_r$ quantifies how effectively its motion appears, to first order, as a common phase shift in the collected reflected field [21,22]. Uniform normal motion gives $\psi = 1$ and $\Gamma_d = \Gamma_r = 1$. Spatial variation, tilt, or opposite-sign lobes reduce the corresponding optical coupling.

Let $F_{\text{m}}(z)$ denote the mechanical restoring force. Quasi-static equilibrium requires $F_{\text{m}}(z)$=$F_{\text{opt}}$. At a given operating point, the incremental force-to-displacement response is described by the differential mechanical compliance $\chi_{\text{m}}(z)$, the inverse local slope of $F_{\text{m}}(z)$. Differentiating the equilibrium relation with respect to $P_\mathrm{d}$ gives the displacement responsivity. Applying the phase relation in Eq. (2) then gives the corresponding phase responsivity:

$$
\begin{aligned}
\chi_m(z) &\equiv \frac{dz}{dF_m} = \left[\frac{dF_m}{dz}\right]^{-1}, \\
\frac{dz}{dP_d} &= \chi_m(z)\frac{dF_\mathrm{opt}}{dP_d} = \frac{\xi}{c}\Gamma_d\chi_m(z), \\
\frac{d\phi}{dP_d} &= \frac{d\phi}{dz}\frac{dz}{dP_d} = \frac{4\pi\xi}{\lambda c}\Gamma_d\Gamma_r\chi_m(z).
\end{aligned}
\tag{3}
$$

Equation (3) makes the dependence on the mechanical operating point explicit. A change in the local slope of the restoring-force curve changes $\chi_\mathrm{m}(z)$, and therefore changes both the displacement and phase responsivities.

To compare devices illuminated by beams of different sizes, we next convert this power responsivity to an intensity-normalized response using the standard nonlinear-optical effective-area convention [23]. For the normalized drive profile defined above, the effective drive-beam area $A_b$, corresponding intensity, and general intensity-normalized response are

$$
\begin{aligned}
A_b &\equiv \frac{\left[\int_S I_d(\mathbf{r})\, dA\right]^2}{\int_S I_d^2(\mathbf{r})\, dA} = \left[\int_S w_d^2(\mathbf{r})\, dA\right]^{-1}, \\
I_b &\equiv \frac{P_d}{A_b}, \\
\frac{d\phi}{dI_b} &= \frac{4\pi\xi}{\lambda c} A_b\Gamma_d\Gamma_r\chi_m(z).
\end{aligned}
\tag{4}
$$

The effective drive-beam area $A_b$ sets the incident-intensity normalization and equals the physical illuminated area for uniform top-hat illumination of the membrane. For arbitrary drive and readout profiles, the sign of $\Gamma_d\Gamma_r$ determines the direction of the phase response. In the same-beam direct-reflection geometry used below, the reflected field is collected in the same spatial mode as the incident beam, so $w_d = w_r = w$ and $\Gamma_d = \Gamma_r = \Gamma$. The overlap product therefore becomes $\Gamma^2 \geq 0$, and we define the nonnegative phase-coupled area as $A_\phi \equiv A_b\Gamma^2$. Because the same optical mode performs both functions, the spatial overlap enters once in driving the motion and once in reading it, producing this squared-overlap factor [21]. Uniform normal motion gives $A_\phi = A_b$. Spatial nonuniformity, mode mismatch, or phase cancellation reduce it. Although $A_\phi$ and $\chi_m(z)$ each depend on the normalization of the modal coordinate, their product, and hence the predicted phase response, is unchanged when both are transformed consistently (Supplementary Section S2). We therefore use its small-signal value, $A_\phi/K$, as the optical comparison metric in Fig. 2(b).

The optical momentum-transfer coefficient $\xi = 2R + A_\mathrm{abs}$ is not included in Fig. 2. It ranges from zero for perfect transmission to two for lossless reflection and determines the force delivered per unit optical power. We leave device-specific $\xi$ out of the literature comparison because reflectance and absorptance were not characterized consistently for those devices. At common wavelength and $\xi$, ratios of $A_\phi/K$ equal ratios of projected small-signal phase slopes with respect to incident intensity. For the trampoline, measured optical coefficients allow the absolute prediction below.

At the rest position $z = 0$, $\chi_m(0) = 1/K$, where $K = (dF_m/dz)_{z=0}$ is the small-signal stiffness. We denote the small-signal phase-coupled compliance by $C_\phi = A_\phi/K$. Writing the same intensity-normalized phase response in the conventional Kerr form defines the device-equivalent nonlinear index

$$\begin{aligned} C_\phi &\equiv A_\phi \chi_{\mathrm{m}}(0) = \frac{A_\phi}{K}, \\ n_{2,\mathrm{eff}}(z) &\equiv \frac{\lambda}{2\pi d}\frac{d\phi}{dI_b} = \frac{2\xi}{cd} A_\phi \chi_{\mathrm{m}}(z), \\ n_{2,\mathrm{eff}}(0) &= \frac{2\xi}{cd} C_\phi. \end{aligned} \tag{5}$$

Because the phase is acquired in reflection, $d$ is an assigned length for reporting the response in conventional nonlinear-optical units, rather than a propagation distance through the membrane. The value of $n_{2,\mathrm{eff}}(z)$ is set by the mechanical compliance, phase-coupled area, optical momentum transfer, and specified interaction length. The zero-displacement value is the small-signal coefficient reported below. The full drive–mechanics–phase derivation and coordinate-normalization analysis are given in Supplementary Sections S1–S2.

Equation (3) applies at any operating point. For a mechanical coordinate with a leading cubic restoring term [24],

$$F_{\mathrm{m}}(z) = Kz + K_3 z^3, \qquad \chi_{\mathrm{m}}(z) = \frac{1}{K + 3K_3 z^2}. \tag{6}$$

For $K_3 > 0$, the response hardens and the differential compliance decreases as $|z|$ increases. The secant compliance $z/F_{\mathrm{m}}(z)$ measures the displacement accumulated under the total applied force. Retained phase responsivity follows the local slope with respect to optical power and is therefore governed by the differential compliance.

Assuming that the optical momentum-transfer coefficient $\xi$, the common drive/readout profile $w(\mathbf{r})$, and its mode-overlap factor $\Gamma$ remain fixed over the excursion, we define $\eta(z)$ as the ratio of the local phase responsivity to its zero-displacement value. Equation (4) then gives

$$\eta(z) \equiv \frac{\chi_{\mathrm{m}}(z)}{\chi_{\mathrm{m}}(0)} = \frac{n_{2,\mathrm{eff}}(z)}{n_{2,\mathrm{eff}}(0)} = \frac{K}{K + 3K_3 z^2}. \tag{7}$$

The linear compliance $1/K$ sets the response at $z = 0$. Equation (7) shows that the departure from this value is governed by the dimensionless combination $3K_3 z^2/K$. Thus, $\sqrt{K/K_3}$ sets the characteristic displacement scale over which the response changes. The local phase responsivity and device-equivalent nonlinear index follow the same retention factor $\eta(z)$.

To quantify the retained range for a chosen tolerance $\delta$, we define $z_\delta$ by $\eta(z_\delta) = 1 - \delta$. For a hardening response with $K_3 > 0$, substituting this condition into Eq. (7) and using the phase relation in Eq. (2) gives

$$\begin{aligned} 1 - \delta &= \eta(z_\delta) = \frac{K}{K + 3K_3 z_\delta^2}, \\ z_\delta &= \left[\frac{K}{3K_3}\frac{\delta}{1-\delta}\right]^{1/2}, \qquad \Phi_\delta = \frac{4\pi\Gamma_r}{\lambda} z_\delta. \end{aligned} \tag{8}$$

In Fig. 2, we set $\delta = 0.01$, so $z_{1\%}$ is the displacement at which the local differential compliance has decreased by 1%. The retained-response range quantifies how far the reflector can move while preserving its local phase responsivity.

### 3. Literature comparison of mechanical resonators for optical phase nonlinearity

To compare mechanical resonators reported in the literature under a common direct-reflection protocol, we compiled data from 38 fundamental or static-dominant devices or modes from 11 independent source datasets reported in 12 papers [20,25–35]. We included only points for which the linear stiffness $K$, cubic stiffness coefficient $K_3$, and the spatial information required for both axes could be reconstructed from traceable source data. The source studies pursued nonlinear material characterization [26], strain engineering [27], nonlinear damping [28,32], linear or noise-referenced dynamic range [29,31], nonlinear mode coupling [30], low-temperature beam characterization [33], stochastic switching [34], optical bistability [20,25], or microwave optomechanical coupling [35]. None of these devices, including the trampoline, was designed to optimize the phase-control metrics introduced here. The inclusion criteria and source-level reconstructions are documented in Supplementary Sections S5–S7.

Panel (a) of Fig. 2 plots small-signal mechanical compliance $1/K$ against the 1%-retained-response displacement $z_{1\%}$. Panel (b) plots phase-coupled compliance $C_\phi = A_\phi/K$ against the same displacement. Both panels therefore use the same retained-range coordinate.

Panel (b) uses a uniform top-hat profile for both drive and readout. The trampoline uses a 40-μm circular aperture on the central pad. The other devices use apertures that follow their selected moving regions. This choice is intentionally generous to those devices because it assumes uniform coupling over these regions and no optical spillover. Their values are geometric projections, not measured optical coupling.

In panel (a), moving upward increases the small-signal mechanical compliance, while moving to the right extends the retained-response range. For the present goal of large phase excursions with little change in responsivity, the upper-right region is favorable. Other applications may instead exploit earlier nonlinear reshaping and therefore favor shorter retained ranges.

In panel (a), the serpentine trampoline [20] ranks third in compliance. Two tapered Si nanowires [31] exceed it in both compliance and retained static range, and most of the 20 nanowires have a longer $z_{1\%}$. Panel (b) asks how much of that mechanical response couples to a reflected phase shift under the common optical projection. Including the phase-coupled area changes the ranking. The trampoline's nearly uniform pad motion over a large optical aperture gives a $C_\phi$ about $2 \times 10^3$ times the next-highest point, while the nanoscale resonators have much smaller phase-coupled areas.

For the trampoline, published optical data allow us to go beyond the common-$\xi$ comparison and calculate its device-specific reflected-phase responsivity.

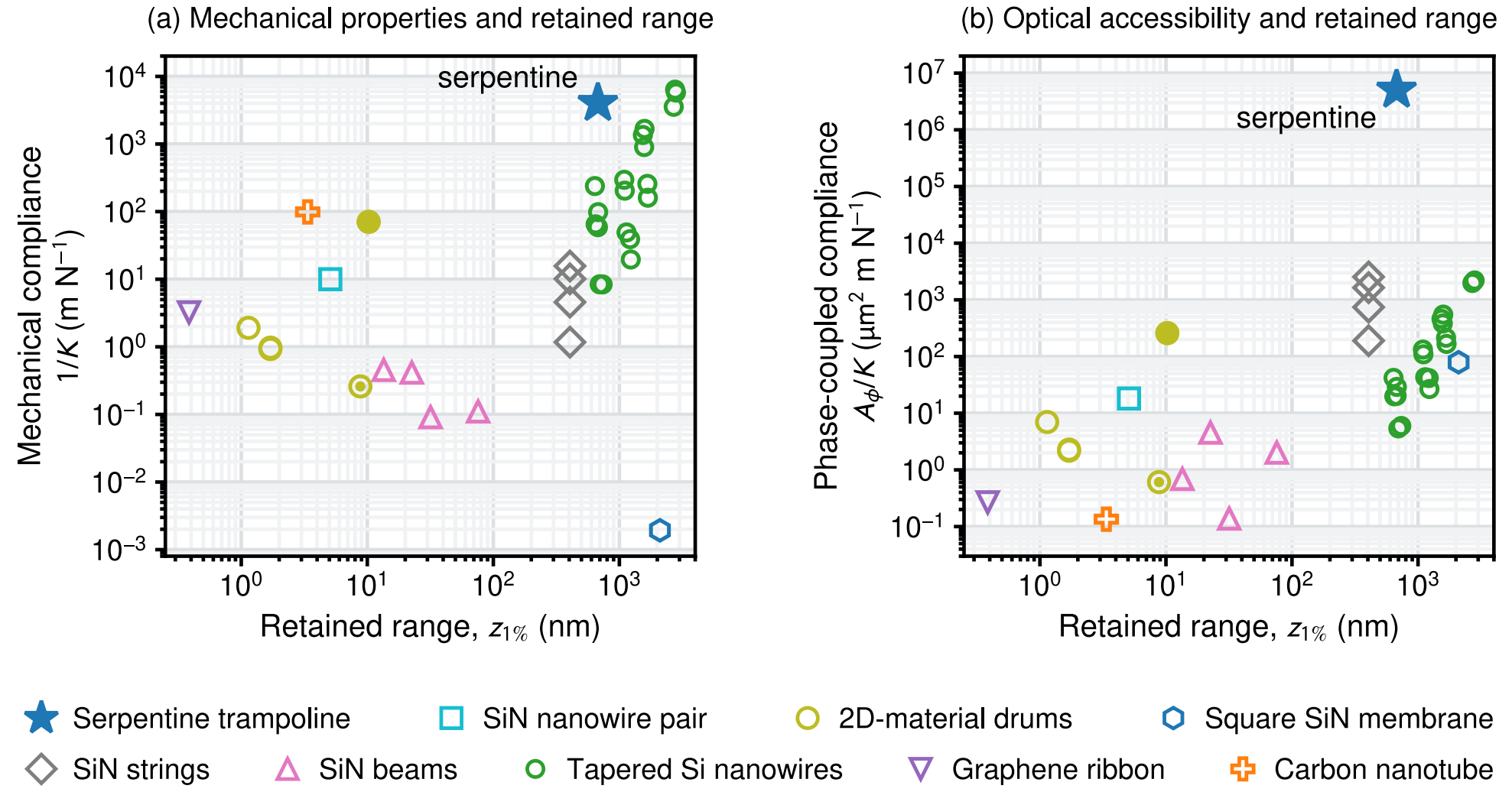

Fig. 2. Literature comparison of mechanical resonators for optical phase nonlinearity. (a) Small-signal mechanical compliance $1/K$ versus the displacement $z_{1\%}$ at which local compliance has decreased by 1%. (b) The same devices and horizontal axis, now with phase-coupled compliance $C_\varphi = A_\varphi/K$ on the vertical axis. The phase-coupled area $A_\varphi$ reflects both the illuminated area and the overlap of mechanical motion with optical drive and reflected-phase readout. The trampoline (filled star) is among the strongest devices mechanically and stands apart once optical accessibility is included. The plotted devices come from Refs. [20,25–35]. Values and sources are listed in Supplementary Table S9, with reconstructions in Supplementary Sections S5–S7.

## 4. Predicted optical phase nonlinearity of the serpentine trampoline

For the trampoline projection, we use the measured mechanics of Ref. [20] and the optical characterization of the closely related Ref. [19] trampoline, as documented in Supplementary Section S4. We adopt a 40-μm-diameter top-hat aperture on the central pad. Static finite-element simulations reported in Supplementary Note 1 of Ref. [19] show approximately 1% displacement variation across the pad. This supports $\Gamma_d \simeq \Gamma_r \simeq 1$ and hence $A_\phi \simeq A_b$. The trampoline's optical absorptance was independently characterized as $A_{\text{abs}} = 4.1 \times 10^{-4}$ [19]. Using $R = 0.40$ [19], $K = 2.48 \times 10^{-4}\,\text{N}\,\text{m}^{-1}$ [20], and $\lambda = 514\,\text{nm}$ [19], Eq. (3) at $z = 0$ gives a low-power phase responsivity of $263\,\text{rad}\,\text{W}^{-1}$. Writing the same response per incident intensity through Eqs. (4) and (5), with the measured pad thickness $d = 50\,\text{nm}$ [19,20], gives $n_{2,\text{eff}}(0) = 5.4 \times 10^{-7}\,\text{m}^2\,\text{W}^{-1}$. The complete source locations and numerical substitutions are given in Supplementary Section S4.

The measured Duffing backbone gives $K_3 = 1.83 \times 10^6\,\text{N}\,\text{m}^{-3}$ for the same central-pad displacement coordinate [20]. The phase relation in Eq. (2) gives $z_{2\pi} = \lambda/(2\Gamma_r) = 257\,\text{nm}$. Substitution into Eq. (7) gives 99.85% retention of the low-power responsivity at this displacement. Equation (8) gives a 1% reduction only at $z_{1\%} = 0.68\,\mu\text{m}$, corresponding to about $5.3\pi$ of accumulated phase. A full $2\pi$ reflected-phase shift therefore lies well within the range over which the measured mechanical response retains its small-signal compliance.

Combining the full restoring force in Eq. (6) with the optical-force relation in Eq. (2) gives approximately 24 mW of incident power for the full $2\pi$ excursion. The explicit finite-excursion calculation is provided in Supplementary Section S4. These predictions isolate the radiation-pressure contribution. Assessing an illuminated device also requires accounting for photothermal forces and temperature-dependent changes in mechanical properties. Figure 3(a) shows the predicted reflected phase versus incident power, while Fig. 3(b) shows the local phase responsivity and $n_{2,\text{eff}}$ across the same phase excursion. Broadband spectral interferometric microscopy provides a direct route for future measurement of the predicted reflected-phase response [36].

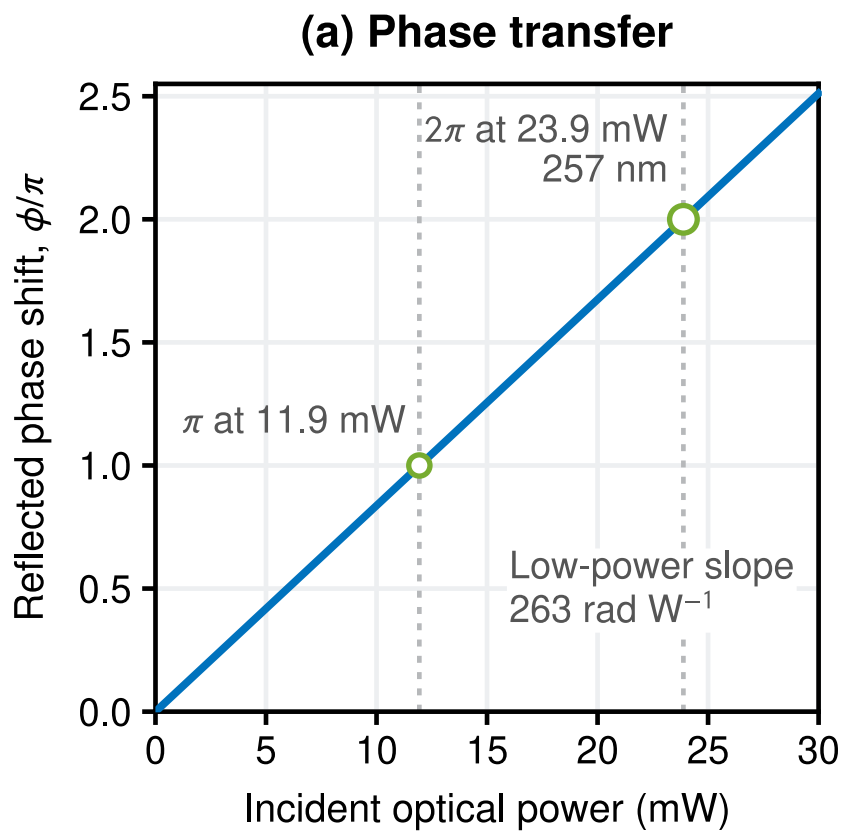

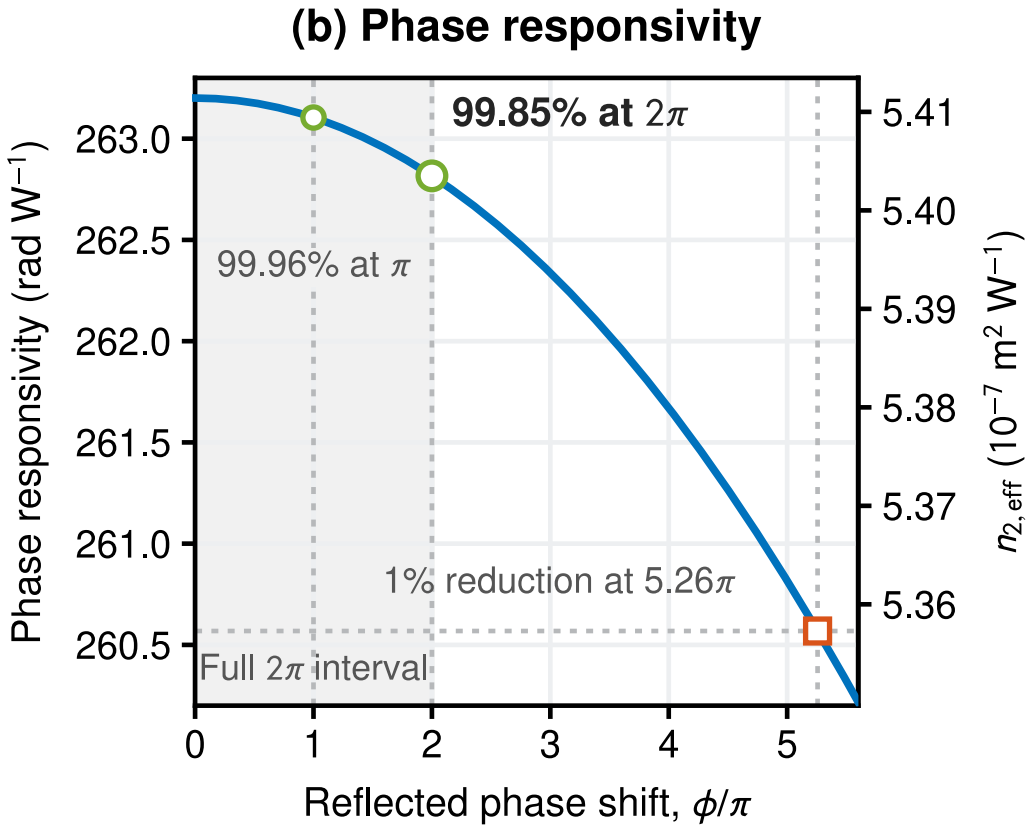

Fig. 3. Predicted optical phase nonlinearity of the serpentine trampoline. Predictions combine previously measured mechanical properties of the trampoline [20] with optical properties measured on a closely related device [19]. (a) Reflected phase shift $\varphi$ versus incident optical power. The small-signal slope is 263 rad $W^{-1}$, and a full $2\pi$ shift requires about 24 mW. (b) Local phase responsivity (left axis) and the corresponding device-equivalent nonlinear index $n_{2,\mathrm{eff}}$ (right axis) versus reflected phase shift $\varphi$. The responsivity retains 99.85% of its small-signal value at $2\pi$ and decreases by 1% only near $5.3\pi$. Source inputs and calculations are given in Supplementary Section S4.

## 5. Discussion and outlook

This work develops a framework for designing compliant reflectors as nonlinear optical phase elements. Its three complementary measures are the small-signal mechanical compliance, the displacement range over which that compliance is retained, and the phase-coupled area that makes the motion accessible to the optical field. Together, they provide a common basis for comparing mechanical strength, usable range, and optical accessibility.

Mechanical strength alone does not determine how effectively a resonator can shift the reflected phase. Several devices have high compliance and a large retained range, and on these criteria the trampoline is among the strongest but is not an outlier. Its broad pad also moves nearly uniformly across the illuminated region, so the same optical field can efficiently drive its motion and acquire a common reflected phase shift. Accounting for this spatial coupling changes the ranking, as shown in Fig. 2(b), where the trampoline is a clear outlier in projected phase-coupled compliance. Its measured restoring force further predicts nearly constant responsivity through a full $2\pi$ reflected-phase shift.

Accessing the projected full-phase range in practice calls for multiphysics co-design of the reflector's optics, mechanics, and thermal pathways. The present optical momentum-transfer coefficient is approximately 0.8. Increasing it toward the lossless-reflection limit of 2 would reduce the incident power required for a given phase shift by up to a factor of 2.5. Suspended photonic-crystal membranes have achieved reflectances approaching 99.99% [37,38]. Peak reflectivity and optical bandwidth are coupled design objectives. The highest-reflectivity membrane demonstrations can be spectrally narrow [38], whereas broadband high-contrast-grating designs combine multiple leaky modes to sustain high reflectivity over fractional bandwidths approaching 30% [39,40]. A purpose-built reflector should therefore optimize the high-reflectivity bandwidth rather than peak reflectivity alone. Lower optical absorption and higher thermal conductance would reduce heating. Measurements on stoichiometric silicon nitride have inferred extinction coefficients of approximately 0.1–1 ppm [41]. Comparative modeling shows that pad and support geometry can substantially modify thermal conductance and relaxation time [42]. The same geometry also changes the linear and cubic mechanical response, so improved thermal transport must be evaluated alongside retained compliance (Supplementary Section S8). Taken together, the cited studies indicate considerable design headroom beyond the reference geometry analyzed here. Response time is another design target. The present high $Q$ reflects the trampoline's original use in weakly damped nonlinear-dynamics experiments [20], and damping can be varied without changing the resonance frequency or Duffing coefficient [43].

The optomechanical phase nonlinearity predicted here is especially timely for cascaded optical processors. These systems call for strong nonlinear operations that can be repeated through many stages without excessive optical loss or changing the frequency of the light passed to the next stage [5,11–13]. A useful phase element should also operate across a practical optical bandwidth and be predictable enough to incorporate into a network model. Our results predict a giant reflected-phase responsivity sustained through a full $2\pi$ shift, using a model based on measured mechanical and optical properties. In the quasi-static regime, radiation pressure imparts this phase response without intrinsic frequency conversion or reliance on a narrow optical resonance. With low-loss, broadband reflectance and suitable dynamics, reflectors designed around the strength–range–aperture metrics identified here could provide a powerful nonlinear phase layer for cascaded optical networks.

Together, these results transform the retrospective comparison into a prospective design strategy. Future devices can jointly engineer mechanical compliance and retained range, phase-coupled area, reflectance, optical loss, thermal pathways, and dynamical response to access large phase shifts with retained responsivity while controlling operating power and response time. The framework therefore establishes a quantitative roadmap toward purpose-built nonlinear phase layers based on compliant reflectors for cascaded optical systems.

### Funding

This work was supported by the Air Force Office of Scientific Research Meta-Imaging MURI grant #FA9550-21-1-0312, the Pritzker Foundation, and the Breakthrough Foundation.

### Acknowledgments

The authors thank Ramon Gao for his work on the fabrication of the serpentine trampoline reported in Ref. [20], for providing the micrograph in Fig. 1(a), and for fruitful discussions.

### Disclosures

The authors declare no conflicts of interest.

### Data Availability

The mechanical and optical parameters used in this work were reported in the cited prior studies. The Supplementary Information provides the source-level inputs, reconstruction methods, Fig. 2 coordinates in Table S9, optical projections, and sensitivity checks.

### Author Contributions

L.M. conceived the study, developed the theoretical and comparative framework, performed the analytical and numerical calculations, compiled the literature dataset, prepared the figures, and wrote the manuscript. H.A.A. supervised the project and contributed to the conceptual framing and interpretation of the results. Both authors discussed the results and revised the manuscript.

See Supplement 1 for supporting content.

# Supplementary Information

# Design of Mechanically Compliant Membrane Reflectors for Giant Optical Phase Nonlinearity

**LIOR MICHAELI[1]* AND HARRY A. ATWATER[2]**

*[1]School of Electrical and Computer Engineering, Faculty of Engineering, Tel Aviv University, 6997801 Tel Aviv, Israel*

*[2]Department of Applied Physics and Materials Science, California Institute of Technology, Pasadena, California 91125, USA*

**Corresponding author: liormic1@tauex.tau.ac.il*

### S1. Drive–mechanics–phase derivation

This section gives the algebra underlying Eqs. (2) and (3) of the main text. Let $S$ be the moving reflective surface and $r$ a position on it. We normalize the spatial mode shape by $\max|\psi(r)| = 1$, so $z$ is the physical maximum displacement of the mode. This spatial convention is distinct from a temporal peak-versus-RMS amplitude convention.

$$\begin{aligned} u_z(\mathbf{r}) &= z\psi(\mathbf{r}), \qquad \max_{\mathbf{r}\in S}|\psi(\mathbf{r})| &&= 1, \\ I_j(\mathbf{r}) &= P_j w_j(\mathbf{r}), \qquad \int_S w_j(\mathbf{r})\,\mathrm{d}A &&= 1, \qquad j \in \{d, r\}. \end{aligned} \tag{S1}$$

The normalized drive and readout profiles integrate to unity. For normal incidence, momentum conservation and $R + T + A_{\text{abs}} = 1$ give the momentum-transfer coefficient $\xi = 1 + R - T = 2R + A_{\text{abs}}$. Projection onto the maximum-displacement mechanical coordinate and the collected optical phase gives

$$\begin{aligned} \Gamma_j &\equiv \int_S w_j(\mathbf{r})\psi(\mathbf{r})\,\mathrm{d}A, \qquad j \in \{d, r\}, \\ F_{\text{opt}} &= \frac{\xi P_d}{c}\Gamma_d, \\ \frac{\mathrm{d}\phi}{\mathrm{d}z} &= \frac{4\pi}{\lambda}\Gamma_r. \end{aligned} \tag{S2}$$

Here $F_{\text{opt}}$ is the generalized optical force conjugate to $z$. Under quasi-static equilibrium, $F_m(z) = F_{\text{opt}}$. Assuming that $\xi$, the optical profiles, and the mode shape remain fixed over the incremental displacement, differentiation with respect to incident drive power gives

$$\frac{\mathrm{d}F_m}{\mathrm{d}z}\frac{\mathrm{d}z}{\mathrm{d}P_d} = \frac{\xi}{c}\Gamma_d. \tag{S3}$$

Define the differential mechanical compliance and small-signal stiffness by

$$\chi_m(z) \equiv \frac{\mathrm{d}z}{\mathrm{d}F_m} = \left(\frac{\mathrm{d}F_m}{\mathrm{d}z}\right)^{-1}, \qquad K \equiv \left.\frac{\mathrm{d}F_m}{\mathrm{d}z}\right|_{z=0}, \qquad \chi_m(0) = \frac{1}{K}. \tag{S4}$$

Combining Eqs. (S2)–(S4) gives

$$\frac{\mathrm{d}\phi}{\mathrm{d}P_d} = \frac{4\pi\xi}{\lambda c}\Gamma_d\Gamma_r\chi_m(z). \tag{S5}$$

### S2. Effective beam area, phase-coupled area, and coordinate normalization

For a fixed drive profile, beam intensity is incident power divided by effective beam area. The standard nonlinear-optical effective area is [2]

$$
\begin{aligned}
A_b &\equiv \frac{\left[\int_S I_d\,(\mathbf{r})\,\mathrm{d}A\right]^2}{\int_S I_d^2\,(\mathbf{r})\,\mathrm{d}A} = \left[\int_S w_d^2\,(\mathbf{r})\,\mathrm{d}A\right]^{-1}, \\
I_b &\equiv \frac{P_d}{A_b}.
\end{aligned} \tag{S6}
$$

Multiplying the power responsivity in Eq. (S5) by the effective beam area gives the intensity-normalized response. In the same-beam direct-reflection geometry used for Fig. 2, drive and readout share the same normalized profile:

$$
\begin{aligned}
&\frac{d\varphi}{dI_b} = A_b \frac{d\varphi}{dP_d} = \frac{4\pi\xi}{\lambda c} A_b \Gamma_d \Gamma_r \chi_m(z), \\
&w_d = w_r = w, \quad \Gamma_d = \Gamma_r = \Gamma, \quad A_\phi \equiv A_b \Gamma^2.
\end{aligned} \tag{S7}
$$

For a uniform top-hat profile over a selected region $\mathcal{A}$, let $A_{\mathrm{ap}}$ denote the area of that region. Then $A_b = A_{\mathrm{ap}}$ and

$$
w(r) = \frac{1}{A_{\mathrm{ap}}}, \quad \Gamma = \frac{1}{A_{\mathrm{ap}}} \int_{\mathcal{A}} \psi(r) dA, \quad A_\phi = A_{\mathrm{ap}} |\Gamma|^2 = \frac{\left|\int_{\mathcal{A}} \psi(r) dA\right|^2}{A_{\mathrm{ap}}}. \tag{S8}
$$

Thus the phase-coupled area is the selected aperture area multiplied by the squared aperture-averaged displacement. The square arises because the same spatial overlap enters once in driving the motion and once in reading the reflected phase. Spatial variation or cancellation reduces the coupling fraction.

Under the consistent rescaling $z' = az, \quad \psi' = \psi/a$, the modal coefficients transform as

$$
\begin{aligned}
&\Gamma'_j = \frac{\Gamma_j}{a}, \quad K' = \frac{K}{a^2}, \quad K_3' = \frac{K_3}{a^4}, \quad A_\phi' = \frac{A_\phi}{a^2}, \\
&z_{1\%}' = a z_{1\%}, \quad \frac{A_\phi'}{K'} = \frac{A_\phi}{K}.
\end{aligned} \tag{S9}
$$

At any displacement, $\chi_m'(z') = a^2 \chi_m(z)$, so $A_\phi' \chi_m'(z') = A_\phi \chi_m(z)$. Its small-signal form is $A_\phi'/K' = A_\phi/K$, as shown in Eq. (S9). All Fig. 2 entries are finally expressed in a maximum-displacement coordinate. A source-specific center, antinode, or tip coordinate is used directly only when the source identifies that location as the displacement maximum. Otherwise, conversion requires a documented mode shape. Entries without a supported conversion to the maximum-displacement coordinate are excluded.

### S3. Retained-compliance range

For the hardening cubic response $F_m(z) = Kz + K_3 z^3$ used for every quantitative point in Fig. 2, $K_3 > 0$ and the differential-compliance retention is

$$
\frac{\chi_m(z)}{\chi_m(0)} = \frac{K}{K + 3K_3 z^2}, \quad z_{1\%} = \sqrt{\frac{K}{3K_3} \frac{0.01}{0.99}}. \tag{S10}
$$

The calculation is performed after conversion to the common maximum-displacement coordinate. This is a differential-compliance retention range, not a noise-referenced resonator dynamic range.

## S4. Trampoline benchmark reconstruction

References [3] and [4] correspond to main-text Refs. [19] and [20], respectively. Ref. [4], Supplementary Table 1, mode 1, reports $k = 2.48 \times 10^{-4}$ N m⁻¹ and $m\alpha = 1.83 \times 10^{6}$ N m⁻³. In the notation used here, $K = k$ and $K_3 = m\alpha$. In that study, we measured the pad motion interferometrically, and Fig. 4(a) shows that the pad lies at an antinode of mode 1. We therefore take the displacement $q$ in its Eq. (1) as the maximum central-pad displacement for that mode. The present-work optical projection adopts a 40-μm-diameter top-hat aperture on the central pad. Ref. [3], Supplementary Note 1 and Fig. S1, reports approximately 1% center-to-edge displacement variation across the pad. We therefore use $\Gamma_d = \Gamma_r = 1$ as the central estimate. The aperture diameter and overlap approximation are present-work choices supported by the source displacement map, not source-measured optical overlaps.

To calculate the absolute response, we combine the mechanical coefficients above with optical quantities characterized for the closely related Ref. [3] trampoline. The operating wavelength is $\lambda = 514$ nm (Supplementary Note 3, text accompanying Fig. S5). At normal incidence, the reflectance is $R = 0.40$ (Supplementary Note 3, Fig. S3), and the absorptance is $A_{\text{abs}} = 4.1 \times 10^{-4}$ (Supplementary Note 3, Eq. (S20) and the following paragraph). We use $d = 50$ nm, the silicon nitride pad thickness stated in the source device descriptions [3,4].

The spring geometries differ between the two devices. The result is therefore a parameter-based projection. Its inputs were not all measured on the same physical device. The complete central-estimate reconstruction is

$$\begin{aligned}
\Gamma_d &= \Gamma_r = 1, \qquad \xi = 2R + A_{\text{abs}} = 0.80041, \\
A_{\text{ap}} &= A_\phi = \pi(20\,\mu\text{m})^2 = 1256.64\,\mu\text{m}^2, \\
C_\phi &= \frac{A_\phi}{K} = 5.067 \times 10^6\,\mu\text{m}^2\,\text{m}\,\text{N}^{-1}, \\
\left.\frac{d\phi}{dP_d}\right|_0 &= \frac{4\pi\xi\Gamma_d\Gamma_r}{\lambda c K} = 263.20\,\text{rad}\,\text{W}^{-1}, \\
n_{2,\text{eff}}(0) &= \frac{2\xi A_\phi}{cdK} = 5.411 \times 10^{-7}\,\text{m}^2\,\text{W}^{-1}, \\
z_{2\pi} &= \frac{\lambda}{2\Gamma_r} = 257.0\,\text{nm}, \\
\eta(z_{2\pi}) &= \frac{K}{K + 3K_3 z_{2\pi}^2} = 0.99854 \;\; (99.854\%), \\
P_{2\pi} &= \frac{c}{\xi\Gamma_d}\left(K z_{2\pi} + K_3 z_{2\pi}^3\right) = 23.884\,\text{mW}, \\
z_{1\%} &= \left[\frac{K}{3K_3}\frac{0.01}{0.99}\right]^{1/2} = 675.5\,\text{nm}, \qquad \Phi_{1\%} = 5.257\pi.
\end{aligned} \tag{S11}$$

Taking $\Gamma_d = \Gamma_r = 0.99$ instead of unity would reduce the small-signal phase responsivity and $n_{2,\text{eff}}$ by 1.99%, increase $z_{2\pi}$ by 1.01%, and increase $P_{2\pi}$ by 2.03%. It would not change the Figure 2 ranking or the conclusions.

## S5. Comparison-map construction protocol

Figure 2 contains 38 devices or modes from 11 independent source datasets reported in 12 papers, spanning nine geometry classes. We include fundamental modes or responses that dominate under static loading. Each entry requires traceable $K$ and $K_3$ in a common maximum-displacement coordinate. Panel (b) also requires a reconstructable displacement profile or an explicit geometric optical projection. The number of points from each study reflects the data it tabulates, not its statistical weight.

Under a static optical load, normal-mode expansion weights each mode by its drive-readout overlap product divided by modal stiffness [1]. Low stiffness contributes only when both overlaps are nonzero, so the fundamental mode need not dominate.

The map retains fundamental or static-dominant responses for which the mechanical coefficients and spatial overlap can be reconstructed. Higher-order resonant entries that lack this information are excluded.

All points use the same uniform top-hat profile for drive and readout. The trampoline uses a 40-μm circular aperture. When a cited study does not define an optical aperture for phase control, we choose an aperture that matches the shape of a physically accessible moving region and assume no spillover. This intentionally favorable convention prevents the comparison from penalizing devices designed for other purposes. It applies only to the optical projection. Unless explicitly stated otherwise, the reconstructed optical quantities are geometric projections, not source-measured reflective phase couplings.

For the mechanical reconstructions, we use source-reported inputs or explicitly cited standard material constants. We do not select parameter extrema to optimize the comparison.

## S6. Comparison-device reconstructions

Each subsection first identifies the source inputs and their exact locations, then explains how we reconstruct the mechanical response and optical projection. We distinguish reported values from analytical or numerical reconstructions, externally adopted constants, and projections introduced in this work. Any sensitivity or qualification that affects the interpretation is discussed alongside the result. Tables S1–S8 collect the source inputs and projections for the individual device families. Table S9 brings together the values used in Fig. 2, rounded for display.

### S6.1 SiN nanowire pair

Section 3 of Ref. [5] reports the restoring law $F(z) = Kz + K_3 z^3$ with $K = 0.10$ N m$^{-1}$ and $K_3 = 1.3 \times 10^{13}$ N m$^{-3}$ for the wider nanowire. The device description identifies 50-nm-thick, 16-μm-long SiN wires, and the thermal model gives a 280-nm width for the wider wire.

For the optical projection, we adopt a rectangular top-hat aperture coincident with the wider wire, $A_{\mathrm{ap}} = Lw$. We approximate the fundamental motion by $\psi(x) = sin(\pi x/L)$, normalized at the midpoint and uniform across the width. This ideal tension-dominated shape is favorable relative to a bending-dominated clamped-clamped mode.

$$\Gamma = (1/L)\int_0^{L} sin(\pi x/L)\, dx = 2/\pi = 0.6366. \qquad \text{(S12)}$$

The adopted aperture is $A_{\mathrm{ap}} = (16)(0.280) = 4.48$ μm². Equation (S8) then gives $A_\phi = 4.48(2/\pi)^2 = 1.816$ μm² and $C_\phi = A_\phi/K = 18.16$ μm² m N$^{-1}$. The retained-compliance coordinate is $z_{1\%} = 5.089$ nm and $1/K = 10.0$ m N$^{-1}$. The 16 μm × 800 nm bounding box of the two-wire pair is not used as a moving aperture.

### S6.2 Circular 2D-material drums

#### Davidovikj et al. [6]:

In Davidovikj et al. [6], source Eq. (1) defines x as the deflection of the membrane center, and source Eq. (4) gives the restoring force in that coordinate. For graphene device 1, *h* = 5 nm and *R* = 2.5 μm appear in the text with Fig. 1(b). The pretension $n_0 = 0.107$ N m$^{-1}$ follows source Eq. (1), and E = 594 GPa follows source Eq. (5).

For $MoS_2$ devices 2 and 3, the Fig. 5 panel annotation gives *h* = 5 nm and *R* = 2.0 μm. The text immediately before Fig. 5 gives E = 315 and 300 GPa and $n_0 = 0.22$ and 0.21 N m$^{-1}$. We adopt ν = 0.165 for graphene [16] and ν = 0.27 for $MoS_2$ [17]. These Poisson ratios are external standard material constants, not values reported by Ref. [6]. Supplementary Note 1 of Ref. [6], Eqs. (11)–(13), gives the reduced-order coefficients below.

$$K = 4.897 n_0, \qquad K_3 = \frac{\pi E h}{R^2(1.269 - 0.967\nu - 0.269\nu^2)}. \qquad \text{(S13)}$$

**Table S1. Source-reported inputs and externally adopted Poisson ratios for Davidovikj et al. [6].**

| Device / material | $n_0$ (N m$^{-1}$) | E (GPa) | *h* (nm) | *R* (µm) | ν |
|---|---|---|---|---|---|
| D1 / graphene | 0.107 | 594 | 5 | 2.5 | 0.165 [16] |
| D2 / $MoS_2$ | 0.22 | 315 | 5 | 2.0 | 0.27 [17] |
| D3 / $MoS_2$ | 0.21 | 300 | 5 | 2.0 | 0.27 [17] |

Supplementary Note 1 of Ref. [6] gives the fundamental mode normalized at the center $\psi(r) = J_0(\alpha_{01}r/R)$, with $\alpha_{01} = 2.40483$, and reports the uniform-force modal overlap $\xi = 0.432$. For a uniform top-hat over the full drum, the same area average reduces to $\Gamma = (2/R^2)\int_0^R J_0(\alpha_{01}r/R)r\,dr$ and evaluates to

$$\Gamma = 2J_1(\alpha_{01})/\alpha_{01} = 0.4318 \approx 0.432. \quad \text{(S14)}$$

We therefore use the full circular aperture $A_{\text{ap}} = \pi R^2$ and $\Gamma_d = \Gamma_r = 0.432$, so that $A_\phi = \pi R^2 \Gamma^2$. This reuses the source mode-shape integral as the present optical drive-readout projection. The resulting phase-coupled areas are 3.664 µm² for D1 and 2.345 µm² for D2 and D3.

Device 4 of Ref. [6] is excluded because the source does not report the $n_0$ required to reconstruct *K*.

### Singh et al. [7,8] and Dolleman et al. [9]:

The two Singh et al. papers [7,8] describe the same multilayer graphene drum. Ref. [7] states this link in the device-description paragraph, where it also reports $f_0$ = 36.233 MHz. The dimensional cubic coefficient $\alpha = 2.3 \times 10^{15}$ N m$^{-3}$ is given around Fig. 4. Supplementary Section I of Ref. [8] gives the 4-µm drum diameter.

Supplementary Section I.B gives a total mass of 0.276 pg and uses that total mass as the effective mass for the displacement calibration. Below, we convert the resulting unit-mean-square source coordinate to the center-displacement coordinate used for the optical projection.

For the single-layer graphene drum reported by Dolleman et al. [9], Main Fig. 1(d) and the accompanying text report $f_0$ = 13.92 MHz. Source Eq. (2) and the following paragraph give an effective modal mass of 1.85 fg from equipartition. Source Eq. (3) and the following text use a center coordinate equal to the physical center deflection divided by the 2.5-µm drum radius. They report a dimensionless cubic coefficient of 200. We denote this coefficient by $\bar{\alpha}$ below to distinguish it from the dimensional coefficient reported by Singh et al.

**Table S2. Source-reported inputs used for the Singh and Dolleman drum reconstructions.**

| Point | Reported input | Value | Exact source location | Role |
|---|---|---|---|---|
| Singh | Drum diameter | 4.0 µm | Ref. [8], SI Section I | Radius and aperture |
| Singh | Total mass | 0.276 pg | Ref. [8], SI Section I.B | Mass in source coordinate |
| Singh | Resonance frequency | 36.233 MHz | Ref. [7], device-description text | Linear stiffness |
| Singh | Dimensional cubic coefficient | $2.3 \times 10^{15}$ N m$^{-3}$ | Ref. [7], text around Fig. 4 | Converted by $\mu^2$ to $K_3$ |
| Dolleman | Effective modal mass | 1.85 fg | Ref. [9], Eq. (2) and following text | Linear stiffness |
| Dolleman | Resonance frequency | 13.92 MHz | Ref. [9], Fig. 1(d) and Eq. (3) text | Linear stiffness |
| Dolleman | Drum radius | 2.5 µm | Ref. [9], text after Eq. (3) | Coordinate scaling and aperture |
| Dolleman | Dimensionless cubic coefficient (source α) | 200 | Ref. [9], Eq. (3) and following text | Cubic-stiffness reconstruction |

For the unit-maximum circular mode in Eq. (S14), the mean-square mode factor is $\mu = \langle\psi^2\rangle$ = 0.2695. The unit-mean-square normalization adopted in Ref. [8] gives $q = \sqrt{\mu}\, z$, where q is the source coordinate and z is center displacement. Transforming the elastic energy multiplies the source linear and cubic stiffnesses by μ and $\mu^2$, respectively, giving

$$K = \mu m(2\pi f_0)^2 = 3.85510\ N\ m^{-1},$$
$$K_3 = \mu^2\alpha = 1.67050 \times 10^{14}\ N\ m^{-3}. \quad \text{(S15)}$$

For Dolleman et al., returning from the radius-normalized source coordinate to the dimensional center-displacement coordinate gives

$$K = m_{\text{eff}}(2\pi f_0)^2 = 0.0141517\ N\ m^{-1},$$
$$K_3 = \bar{\alpha}K/R^2 = 4.52856 \times 10^{11}\ N\ m^{-3}, \quad \bar{\alpha} \equiv R^2K_3/K = 200. \quad \text{(S16)}$$

Both points use the full-drum optical projection defined by Eq. (S14). The aperture covers the full circular moving area, and the drive and readout overlaps are both 0.432. The resulting phase-coupled areas are 2.345 μm² for Singh and 3.664 μm² for Dolleman.

### S6.3 Strain-engineered SiN strings

Li et al. [10] study four devices with a common central SiN string and different support lengths. The Results text gives $L$ = 200 μm and $w$ = 2 μm. Figure 1(e) identifies $w_s$ = 1 μm, θ = 0, and $L_s$ = 30, 70, 110, and 150 μm. It also reports the corresponding measured mass-normalized Duffing coefficients β = 8.28, 2.12, 0.96, and 0.62 × $10^{22}$ $m^{-2}$ $s^{-2}$.

The text after source Eq. (1) defines q as the displacement of the string center. Source Supplementary Note 2 assumes u_z(x,$t$) = q($t$) sin(πx/$L$), so q is also the maximum displacement of this symmetric fundamental mode.

**Table S3. Source-reported inputs used for the Li et al. [10] string reconstruction.**

| Input | Value | Exact source location | Role |
|---|---|---|---|
| Central-string geometry | $L$ = 200 μm, $w$ = 2 μm | Results, opening paragraph | Mass, mode, aperture |
| Thickness | $h$ = 340 nm | Methods, Sample fabrication | Effective mass |
| Material properties | ρ = 3100 kg $m^{-3}$, $\sigma_0$ = 1.06 GPa, E = 271 GPa, ν = 0.23 | Methods, final paragraph | Mechanical model |
| Support geometry | $w_s$ = 1 μm, θ = 0, $L_s$ = 30, 70, 110, 150 μm | Fig. 1(e) caption | Device identity |
| Measured β | (8.28, 2.12, 0.96, 0.62) × $10^{22}$ $m^{-2}$ $s^{-2}$ | Fig. 1(e) caption | Cubic stiffness |
| Coordinate and mode | q, with u_z = q sin(πx/$L$) | Text after Eq. (1) and source Supplementary Note 2, Eq. (S2) | Center/maximum displacement |

Source Eqs. (3) and (4) and Supplementary Eqs. (S12) and (S13) use the same support-tuning factor T in α and β. With A = hw, the dimensional coefficients are therefore

$$m_{\text{eff}} = \rho hwL/2, \quad K_3 = m_{\text{eff}}\beta, \quad K = m_{\text{eff}}\alpha,$$
$$\alpha = \frac{\pi^2(1-\nu)\sigma_0 T}{\rho L^2}, \quad \beta = \frac{\pi^4 ET}{4\rho L^4}, \quad T = (1 + \frac{2EA}{k_{\text{in}}L})^{-1}. \quad \text{(S17)}$$

Because T is common to α and β, it cancels from their ratio. Thus the reported β values can be combined with the source analytical model without separately reconstructing the support stiffness:

$$\frac{K}{K_3} = \frac{\alpha}{\beta} = \frac{4(1-\nu)\sigma_0 L^2}{\pi^2 E}. \quad \text{(S18)}$$

Substitution of the common source inputs gives

$$m_{\text{eff}} = 2.108 \times 10^{-13}\ kg, \quad K/K_3 = 4.88256 \times 10^{-11}\ m^2. \quad \text{(S19)}$$

The resulting model-assisted reconstruction is reported in consolidated Table S9. The directly reported inputs are the geometry, material constants, coordinate, and β. The four $K$ values are inferred from the source model and are not reported as four direct stiffness measurements.

For a hardening cubic response and a fractional differential-compliance loss δ, the exact retained-range expression is

$$z_{\delta} = \sqrt{\frac{K}{3K_3}\frac{\delta}{1-\delta}}, \quad z_{1\%} = 405.458 \text{ nm}. \tag{S20}$$

The identical $K/K_3$ ratio therefore gives the same $z_{1\%}$ for all four support lengths. The exact factor 0.01/0.99 is retained rather than replaced by the approximation 0.01.

For the present-work uniform top-hat over the projected rectangle of the central string, the source sinusoidal mode gives

$$A_{\mathrm{ap}} = Lw = 400 \ \mu\mathrm{m}^2, \\ \Gamma = \frac{1}{L}\int_0^L sin(\pi x/L)dx = \frac{2}{\pi}, \quad A_{\phi} = A_{\mathrm{ap}}\Gamma^2 = 162.114 \ \mu\mathrm{m}^2. \tag{S21}$$

Li et al. read the string center with a laser Doppler vibrometer rather than the full-aperture optical average adopted here.

We can check the sensitivity to the measured frequency for the $L_s$ = 150 µm device. It is the only one of the four for which the source gives a numerical measured $f_0$. The source reports measured $f_0$ = 103 kHz and β = 6.20 × $10^{21}$ $m^{-2}$ $s^{-2}$, compared with analytical values of 94 kHz and 7.21 × $10^{21}$ $m^{-2}$ $s^{-2}$.

Using the measured frequency directly gives $K$ = 0.0882886 N $m^{-1}$, $1/K$ = 11.3265 m $N^{-1}$, $z_{1\%}$ = 476.917 nm, and $C\varphi$ = 1836.18 µm² m $N^{-1}$. The common model route used in Fig. 2 instead gives $C\varphi$ = 2540.45 µm² m $N^{-1}$, which favors the Li comparison point. Under the measured-frequency alternative, the largest Molina point becomes the next-highest optical point. The trampoline-to-next-point ratio then increases from 1995 to approximately 2304.

### S6.4 SiN membrane and beam resonators

This subsection reconstructs the five retained SiN membrane and beam points from source-reported mechanical inputs. Table S4 records the exact source locations. The optical apertures and overlaps are present-work top-hat projections under the common convention of Section S5.

**Table S4. Source-reported inputs used for the Das, Ma, and Defoort reconstructions.**

| Point | Reported input | Value | Exact source location and role |
|---|---|---|---|
| Das (1,1) | Geometry and density | $L$ = 500 µm, $h$ = 100 nm, ρ = 3170 kg $m^{-3}$ | Fig. 1(b) caption and device paragraph for $L$ and $h$. Supplementary Material parameter list for ρ. Used in m_eff. |
| Das (1,1) | Frequency and Duffing coefficient | $f_{11}$ = 814 kHz, $\alpha_{11}$ = 2.0 × $10^{22}$ $m^{-2}$ $s^{-2}$ | Main text with Fig. 2(b) for $f_{11}$. Text following source Eq. (6) and Fig. 2(d) for $\alpha_{11}$. |
| Das (1,1) | Coordinate and mode | $w$ = ψz, ψ = sin(πx/$L$)sin(πy/$L$) | Source Eq. (3) and preceding mode expansion. Figure 2(c) and source Eq. (6) identify z as the maximum displacement amplitude. |
| Ma, $L$ = 50 µm | Fundamental-mode row | $w$ = 900 nm, $h$ = 100 nm, f = 4.954 MHz, $K$ = 9.05 N $m^{-1}$, α = 5.69 × $10^{-4}$ MHz² $nm^{-2}$ | Source Table I, nominal 50 µm × 900 nm × 100 nm beam, n = 1 row. |
| Ma, $L$ = 30 µm | Fundamental-mode row | $w$ = 900 nm, $h$ = 100 nm, f = 3.488 MHz, $K$ = 2.41 N $m^{-1}$, α = 3.18 × $10^{-3}$ MHz² $nm^{-2}$ | Source Table I, nominal 30 µm × 900 nm × 100 nm beam, n = 1 row. |
| Ma beams | Coordinate and mass relation | Antinode displacement. Effective mass reconstructed from $K$ and f. | Source Eq. (6) and the text after Eq. (7). The source explicitly calls the modal coordinate the temporal peak displacement at the antinodes. |

| Point | Reported input | Value | Exact source location and role |
|---|---|---|---|
| Defoort A1 | Frequency, stiffness, backbone slope | $f_0$ = 7 MHz, $K$ = 2.2 N m$^{-1}$, RMS slope = 9.7 × 10$^{19}$ Hz m$^{-2}$ | Results paragraph for $f_0$. Fig. 3 fit annotation for $K$. Fig. 4 quadratic fit for the backbone slope. |
| Defoort A2 | Frequency, stiffness, backbone slope | $f_0$ = 14 MHz, $K$ = 11 N m$^{-1}$, RMS slope = 3.5 × 10$^{19}$ Hz m$^{-2}$ | Results paragraph for $f_0$. Fig. 3 fit annotation for $K$. Fig. 4 quadratic fit for the backbone slope. |
| Defoort beams | Representative geometry | $L$ = 15 µm, $w$ = 250 nm, $t$ = 100 nm, with 30-nm Al coating | Source Fig. 1 caption. The source calls this a typical beam and states that A1 and A2 have similar dimensions. Used for both optical projections. |
| Defoort beams | Temporal amplitude convention | Source displacement is RMS | Defoort states that force and displacement are calibrated through its Ref. 15. That calibration reports experimental results as RMS [18]. Converted to temporal peak amplitude in Eq. (S24). |

**Das et al. [13]:**

For the fundamental square-membrane mode reported by Das et al. [13], the mass-normalized equation and center-normalized sine-product mode give

$$m_{\text{eff}} = \rho h L^2/4, \quad K = m_{\text{eff}}(2\pi f_{11})^2, \quad K_3 = m_{\text{eff}}\alpha_{11}, \\ A_{\text{ap}} = L^2, \quad \Gamma = 4/\pi^2, \quad A_\phi = A_{\text{ap}}\Gamma^2 = 16L^2/\pi^4. \tag{S22}$$

Only the (1,1) mode is retained. It is the fundamental response and has nonzero uniform-drive and uniform-readout overlap over the full square.

**Ma et al. [11]:**

For the fundamental modes of the two beams reported by Ma et al. [11], the source Table I coefficient is mass-normalized. Since 1 MHz$^2$ nm$^{-2}$ = 10$^{30}$ s$^{-2}$ m$^{-2}$, the dimensional mechanical and present-work optical coordinates are

$$m_{\text{eff}} = K/(2\pi f)^2, \quad \alpha_{\text{SI}} = 10^{30}\alpha_{\text{table}}, \quad K_3 = m_{\text{eff}}\alpha_{\text{SI}}, \\ A_{\text{ap}} = Lw, \quad \Gamma = 2/\pi, \quad A_\phi = 4Lw/\pi^2. \tag{S23}$$

Only n = 1 from each beam is retained as a separate device point. The higher rows of Table I are resonant modes of the same devices rather than additional device configurations. Under the adopted uniform same-beam static projection, even modes cancel and the overlap of higher odd modes decreases while their stiffness increases.

**Defoort et al. [12]:**

For Defoort et al. [12], the measured quadratic backbone is written in the source-calibration RMS coordinate. Converting it to the common temporal peak displacement and comparing it with the Duffing backbone gives

$$z = \sqrt{2}x_{\text{RMS}}, \quad f = f_0 + b_{\text{RMS}}{x_{\text{RMS}}}^2 = f_0 + (b_{\text{RMS}}/2)z^2, \\ \frac{\Delta f}{f_0} = \frac{3K_3 z^2}{8K}, \quad K_3 = \frac{4K b_{\text{RMS}}}{3f_0}. \tag{S24}$$

For both Defoort devices, the present-work optical projection uses the typical 15 µm × 250 nm projected rectangle from the source Fig. 1 caption with $\Gamma = 2/\pi$. The source reports two samples of similar dimensions but does not tabulate a separate exact geometry for A1 and A2. The same representative area is therefore applied to both and is identified as an assumption.

**S6.5 Graphene ribbon and carbon nanotube resonators**

This subsection reconstructs the two retained Eichler points while separating the source total-mass scaling from the generalized force coordinate used throughout this work. Table S5 records every source input and its exact location. The source displacement $x_0$ is the temporal peak amplitude at the resonator midpoint, so no peak-to-RMS conversion is required.

**Table S5. Source-reported inputs used for the Eichler reconstructions.**

| Point | Reported input | Value | Exact source location and role |
|---|---|---|---|
| Graphene ribbon | Geometry | $L$ = 1.7 μm, $w$ = 120 nm | Main Fig. 3 caption. Defines the projected moving rectangle. |
| Graphene ribbon | Total suspended mass | $m_{tot} = 3.9 \times 10^{-19}$ kg | Supplementary Section E, paragraph accompanying Fig. S8. Used in the linear stiffness reconstruction. |
| Graphene ribbon | Operating frequency | $f_0 \approx 200.2$ MHz | Read from Main Fig. 3(c,d). This is a plot-derived value, not a text-tabulated number. |
| Graphene ribbon | Source-scaled cubic coefficient | $\alpha_{src} = 1.4 \times 10^{16}$ N m$^{-3}$ | Main Fig. 3 caption, improved fit including finite linear damping. The $\gamma = 0$ fit gives $1.9 \times 10^{16}$ N m$^{-3}$. |
| Carbon nanotube | Geometry | $L$ = 840 nm, $r$ = 2 nm | Main Fig. 2 caption. Supplementary Section E states that $L$ and r were measured by AFM. The projected width is $w = 2r = 4$ nm. |
| Carbon nanotube | Total suspended mass | $m_{tot} = 7.87 \times 10^{-21}$ kg | Supplementary Section E, text following source Eq. (S17). Used in the linear stiffness reconstruction. |
| Carbon nanotube | Operating frequency | $f_0 \approx 254.9$ MHz | Read from Main Fig. 2(b). This is a plot-derived value, not a text-tabulated number. |
| Carbon nanotube | Source-scaled cubic coefficient | $\alpha_{src} = 6.0 \times 10^{12}$ N m$^{-3}$ | Main Fig. 2 caption for the tensile-nanotube resonance-shift fit. |
| Both devices | Coordinate and equation scaling | $x_0$ is the maximum midpoint amplitude. $m_{tot}$ is the total suspended mass. | Supplementary Section H immediately after source Eqs. (S20a)–(S20b). Source Eq. (S21) gives the associated force-shape factor. |

Eichler et al. scale their reduced equation so that its mass is the total suspended mass. For the tensioned fundamental mode, the source force-shape factor is $g = \langle\psi\rangle/\langle\psi^2\rangle = 4/\pi$. This corresponds to the unit-maximum sine mode below. Returning to the generalized force coordinate used in Sections S1–S2 gives

$$\psi(x) = sin(\pi x/L), \quad I_1 = \langle\psi\rangle = 2/\pi, \quad I_2 = \langle\psi^2\rangle = 1/2,$$
$$m_{\text{eff}} = I_2 m_{\text{tot}} = m_{\text{tot}}/2, \quad K = I_2 m_{\text{tot}}(2\pi f_0)^2, \quad K_3 = I_2\alpha_{src} = \alpha_{src}/2. \tag{S25}$$

The common factor $I_2$ therefore leaves $K/K_3$ and $z_1\%$ unchanged, while the generalized compliance $1/K$ is twice the value obtained by inserting the source-scaled stiffness directly.
For the present-work uniform top-hat matched to the projected moving rectangle, the same sine mode gives

$$w_{CNT} = 2r, \quad A_{\text{ap}} = Lw, \quad \Gamma_d = \Gamma_r = I_1 = 2/\pi,$$
$$A_\phi = A_{\text{ap}}\Gamma_d\Gamma_r = 4Lw/\pi^2. \tag{S26}$$

The no-spillover qualification is especially important for the 4-nm-wide nanotube projection. These optical entries are therefore numerical present-work projections rather than source-measured optical couplings.

Using the alternate graphene fit $\alpha_{src} = 1.9 \times 10^{16}$ N m$^{-3}$ gives $K_3 = 9.5 \times 10^{15}$ N m$^{-3}$ and $z_{1\%} =$ 0.33069 nm. $K$, $A\varphi$, and $C\varphi$ are unchanged. Because both operating frequencies are read from plotted traces, they are retained only to plot-level precision. Rounding them to 200 and 255 MHz changes $K$ and $C\varphi$ by at most 0.20% and does not change any ranking.

### S6.6 Tapered Si nanowires

For Molina et al. [14], Main Table 1 (Nano Letters p. 6621) reports 20 devices. It tabulates their lengths, base diameters, taper coefficients, effective masses, quality factors, experimental and theoretical frequencies, and dynamic ranges. Table S6 transcribes the inputs used in Fig. 2. The quality factors, theoretical frequencies, and dynamic ranges are used only for the numerical cross-check below, so we do not reproduce them in a separate table.

The reconstruction follows the source SI. Equation (S2) defines the unit-maximum displacement coordinate, and Eqs. (S5)–(S6) give the singly clamped equation and its coefficients. Equations (S15)–(S16) separate geometric and inertial nonlinearities. The linear taper is introduced in Eqs. (S36)–(S38) and source Table S4. Source Eq. (S32) provides the dynamic-range cross-check.

**Table S6. Source-reported Molina inputs used in the Fig. 2 reconstruction.**

| Device | $L$ (µm) | $D_0$ (nm) | $\alpha_T$ | $m_{eff}$ (fg) | $f_{exp}$ (kHz) |
|---|---|---|---|---|---|
| 1 | 43.2 | 241 | 0.89 | 46 | 294 |
| 2 | 44.2 | 256 | 0.88 | 56 | 280 |
| 3 | 43.7 | 276 | 0.81 | 100 | 267 |
| 4 | 25.9 | 199 | 0.72 | 50 | 550 |
| 5 | 17.6 | 212 | 0.34 | 145 | 925 |
| 6 | 25.7 | 237 | 0.69 | 81 | 589 |
| 7 | 25.1 | 219 | 0.73 | 55 | 582 |
| 8 | 27 | 294 | 0.52 | 255 | 621 |
| 9 | 27.1 | 326 | 0.46 | 378 | 646 |
| 10 | 17.5 | 190 | 0.38 | 105 | 903 |
| 11 | 18.1 | 309 | 0.28 | 369 | 1181 |
| 12 | 19.3 | 347 | 0.27 | 502 | 1135 |
| 13 | 10.3 | 145 | 0.46 | 29 | 1912 |
| 14 | 10.3 | 192 | 0.29 | 79 | 2222 |
| 15 | 11.6 | 338 | 0.17 | 357 | 2909 |
| 16 | 19.5 | 422 | 0.24 | 802 | 1270 |
| 17 | 10.8 | 183 | 0.31 | 71 | 1906 |
| 18 | 10.7 | 194 | 0.24 | 94 | 2139 |
| 19 | 10.5 | 203 | 0.29 | 91 | 2134 |
| 20 | 11 | 326 | 0.18 | 308 | 3127 |

$$m\ddot{z} + \frac{m\omega_0}{Q}\dot{z} + Kz + \frac{\beta_G}{L^2}z^3 + \frac{\beta_I}{L^2}(z\dot{z}^2 + z^2\ddot{z}) = F_0\cos\omega t. \tag{S27}$$

Here z is the nanowire-tip displacement and $\phi(1) = 1$. The symbol m denotes the source-tabulated effective mass in this same reduced-order coordinate. We use the tabulated value directly rather than recalculating it from the scalar diameter. The linear stiffness uses the experimental frequency, $K = m(2\pi f_exp)^2$. The theoretical frequency, $Q$, and dynamic range do not enter the Fig. 2 point.

$$\begin{aligned} K_{3,\mathrm{static}} &= \frac{\beta_G}{L^2} = \frac{K\alpha_G}{L^2}, \qquad \alpha_G \equiv \frac{\beta_G}{K}, \\ \alpha_I &\equiv \frac{2\beta_I}{3m}, \\ \alpha_{\mathrm{NL}} &= \alpha_G - \alpha_I, \\ K_{3,\mathrm{dynamic}} &= \frac{K\alpha_{\mathrm{NL}}}{L^2}. \end{aligned} \tag{S28}$$

We reconstruct the taper-dependent mode and dimensionless coefficients from the source equations below. Molina et al. do not tabulate the geometric coefficient $\alpha G$, so we obtain it by numerically solving the tapered Euler–Bernoulli eigenproblem.

$$
\begin{aligned}
m &= \rho L S_0 \int_0^1 (1-\alpha_T s)^2\, \varphi^2(s)\, \mathrm{d}s, \\
K &= \frac{EI_0}{L^3} \int_0^1 (1-\alpha_T s)^4\, [\varphi''(s)]^2\, \mathrm{d}s, \\
\beta_G &= \frac{2EI_0}{L^3} \int_0^1 (1-\alpha_T s)^4\, [\varphi'(s)\varphi''(s)]^2\, \mathrm{d}s, \\
\beta_I &= \rho L S_0 \int_0^1 (1-\alpha_T s)^2 \left[\int_0^s [\varphi'(t)]^2\, \mathrm{d}t\right]^2 \mathrm{d}s.
\end{aligned}
\tag{S29}
$$

The nonlinear-inertia contribution vanishes when velocity and acceleration are zero. Figure 2 therefore uses the static geometric coefficient $K_3$,static = $K\alpha$G/$L^2$. The cancellation-reduced resonant coefficient $K_3$,dynamic = $K\alpha$NL/$L^2$ remains an audit quantity. Across the 20 devices, $K_3$,static/$K_3$,dynamic ranges from 6.91 to 13.99, so the static choice is more restrictive for the nanowires than the resonant coefficient.
The source experiment detected vibration-induced backscatter from a 633-nm beam with an approximately 4-µm waist diameter. It did not measure reflective phase from a uniform aperture. For Fig. 2, we construct a favorable optical bound by assuming uniform illumination with no spillover over either the full projected wire or a contiguous region near the tip. We treat the reported scalar diameter D(s) as the projected width. This is a geometric assumption: the source describes a hexagonal base evolving toward a polyhedral tip and does not report an optical orientation. The corresponding areas are

$$
\begin{aligned}
A_{\phi,\mathrm{full}} &= L D_0 \frac{\left[\int_0^1 (1-\alpha_T s)\, \varphi(s)\, \mathrm{d}s\right]^2}{\int_0^1 (1-\alpha_T s)\ \mathrm{d}s}, \\
A_{\phi,\mathrm{sel}} &= \max_{0\le s_0<1} L D_0 \frac{\left[\int_{s_0}^1 (1-\alpha_T s)\, \varphi(s)\, \mathrm{d}s\right]^2}{\int_{s_0}^1 (1-\alpha_T s)\ \mathrm{d}s}.
\end{aligned}
\tag{S30}
$$

We numerically solve the tapered Euler–Bernoulli problem and evaluate the equations above. Refining the mesh from 40 to 80 elements changes $\alpha G$, $\alpha NL$, and the selected-aperture factor by less than $3 \times 10^{-6}$ in relative terms. The largest discrepancy between the numerical $\alpha NL$ and the source Table S4 polynomial is 2.85%. This reflects the accuracy of the published fit rather than mesh error.

We also check the reconstruction against the source dynamic ranges. Using the source-reported $Q$ and theoretical frequency in source Eq. (S32), the reconstructed dynamic ranges differ from the rounded theoretical values in source Table 1 by at most 0.624 dB. The inputs used for this cross-check do not enter the Fig. 2 points.

The selected interval begins at $s_0$ = 0.5006–0.5654 and covers the distal 43.5–49.9% of each wire. For each device, we choose the interval that maximizes $A\varphi$ under the stated no-spillover construction. This increases $A\varphi$ by a factor of 1.49–2.15 relative to uniform illumination of the full projected side. The mechanical quantities are numerically reconstructed from source inputs. The optical quantities are projections introduced in this work.

**Table S7. Reconstructed static Molina mechanical quantities used in Fig. 2.**

| Device | $K$ (N m$^{-1}$) | $\alpha_G$ | $K_{3,\mathrm{static}}$ (N m$^{-3}$) | $z_{1\%}$ (nm) |
|---|---|---|---|---|
| 1 | $1.5697 \times 10^{-4}$ | 0.81973 | $6.8947 \times 10^{4}$ | 2768.7 |
| 2 | $1.7333 \times 10^{-4}$ | 0.83178 | $7.3796 \times 10^{4}$ | 2812.2 |

| Device | $K$ (N m$^{-1}$) | $\alpha_G$ | $K_{3,static}$ (N m$^{-3}$) | $z_{1\%}$ (nm) |
|---|---|---|---|---|
| 3 | $2.8144 \times 10^{-4}$ | 0.88267 | $1.3008 \times 10^{5}$ | 2699.0 |
| 4 | $5.9711 \times 10^{-4}$ | 0.89965 | $8.0081 \times 10^{5}$ | 1584.5 |
| 5 | 0.0048979 | 0.85662 | $1.3545 \times 10^{7}$ | 1103.4 |
| 6 | 0.0011094 | 0.89965 | $1.5111 \times 10^{6}$ | 1572.2 |
| 7 | $7.3548 \times 10^{-4}$ | 0.89921 | $1.0497 \times 10^{6}$ | 1535.9 |
| 8 | 0.0038822 | 0.88194 | $4.6967 \times 10^{6}$ | 1668.3 |
| 9 | 0.0062275 | 0.87341 | $7.4062 \times 10^{6}$ | 1682.6 |
| 10 | 0.0033801 | 0.86209 | $9.5148 \times 10^{6}$ | 1093.7 |
| 11 | 0.020318 | 0.84875 | $5.2639 \times 10^{7}$ | 1140.0 |
| 12 | 0.02553 | 0.84748 | $5.8085 \times 10^{7}$ | 1216.5 |
| 13 | 0.0041854 | 0.87341 | $3.4457 \times 10^{7}$ | 639.5 |
| 14 | 0.015398 | 0.85003 | $1.2338 \times 10^{8}$ | 648.3 |
| 15 | 0.11927 | 0.83546 | $7.405 \times 10^{8}$ | 736.4 |
| 16 | 0.051067 | 0.84374 | $1.1331 \times 10^{8}$ | 1231.8 |
| 17 | 0.010183 | 0.85263 | $7.4435 \times 10^{7}$ | 678.7 |
| 18 | 0.016979 | 0.84374 | $1.2513 \times 10^{8}$ | 675.9 |
| 19 | 0.01636 | 0.85003 | $1.2614 \times 10^{8}$ | 660.8 |
| 20 | 0.1189 | 0.83661 | $8.2206 \times 10^{8}$ | 697.8 |

**Table S8. Present-work full-side and selected-region optical projections for the Molina devices.**

| Device | $s_0$ | $A_{ap,sel}$ (µm²) | $A_{\phi,full}$ (µm²) | $A_{\phi,sel}$ (µm²) | $C_{\phi,full}$ (µm² m N$^{-1}$) | $C_{\phi,sel}$ (µm² m N$^{-1}$) |
|---|---|---|---|---|---|---|
| 1 | 0.5652 | 1.3738 | 0.1553 | 0.3330 | 989.09 | 2121.7 |
| 2 | 0.5654 | 1.5304 | 0.1789 | 0.3812 | 1032.4 | 2199.1 |
| 3 | 0.5623 | 1.9392 | 0.2761 | 0.5580 | 980.89 | 1982.5 |
| 4 | 0.5531 | 1.0154 | 0.1732 | 0.3280 | 290.11 | 549.35 |
| 5 | 0.5139 | 1.3470 | 0.3419 | 0.5359 | 69.799 | 109.42 |
| 6 | 0.5497 | 1.2763 | 0.2285 | 0.4242 | 206 | 382.37 |
| 7 | 0.5543 | 1.0602 | 0.1778 | 0.3390 | 241.71 | 460.86 |
| 8 | 0.5309 | 2.2417 | 0.4933 | 0.8325 | 127.06 | 214.44 |
| 9 | 0.5248 | 2.7257 | 0.6330 | 1.0396 | 101.64 | 166.93 |
| 10 | 0.5174 | 1.1421 | 0.2821 | 0.4486 | 83.453 | 132.73 |
| 11 | 0.5089 | 2.1664 | 0.5706 | 0.8767 | 28.084 | 43.15 |
| 12 | 0.5081 | 2.6235 | 0.6950 | 1.0645 | 27.224 | 41.697 |
| 13 | 0.5248 | 0.4608 | 0.1070 | 0.1757 | 25.566 | 41.989 |
| 14 | 0.5097 | 0.7573 | 0.1983 | 0.3057 | 12.878 | 19.85 |
| 15 | 0.5006 | 1.7082 | 0.4771 | 0.7097 | 4.0001 | 5.9509 |
| 16 | 0.5058 | 3.3320 | 0.8977 | 1.3623 | 17.579 | 26.677 |
| 17 | 0.5113 | 0.7396 | 0.1913 | 0.2968 | 18.787 | 29.15 |
| 18 | 0.5058 | 0.8405 | 0.2264 | 0.3437 | 13.337 | 20.24 |
| 19 | 0.5097 | 0.8163 | 0.2137 | 0.3295 | 13.064 | 20.137 |
| 20 | 0.5013 | 1.5466 | 0.4298 | 0.6412 | 3.6152 | 5.3929 |

For device 2, $K = 1.7333 \times 10^{-4}$ N m$^{-1}$ and $\alpha G = 0.83178$ give $K_3$,static = $7.3796 \times 10^{4}$ N m$^{-3}$ and $z_1$% = 2812.2 nm. This is also the largest Molina point in Fig. 2(b). Its selected region gives $C\varphi$ = 2199.1 µm² m N$^{-1}$, while full-side illumination gives 1032.4 µm² m N$^{-1}$.

With the Li model reconstruction used in the main map, the 150-μm-support SiN string is the next-highest point, and the trampoline-to-next-point ratio is 1995. Using the measured Li frequency instead makes Molina device 2 the next-highest point and increases the ratio to approximately 2304.

The $z_1$% values of the compared resonators are mechanical ranges. A finite optical phase range also depends on the mode shape and illumination. The explicit 2π prediction is therefore made only for the nearly uniform trampoline pad.

### S7. Consolidated point-level results and reproducibility

Sections S6.1–S6.6 explain how all Fig. 2 points are reconstructed, including the exact source locations and aperture models. Table S9 collects the resulting values, rounded for display. The Fig. 2(a) ordinate is $1/K$, obtained directly from the tabulated $K$. The calculations retain full numerical precision and check dimensional consistency, $z_1$% by back-substitution, coordinate invariance, source reconstructions, tapered-nanowire convergence, and ranking sensitivity.

The first sensitivity test reduces both the trampoline aperture diameter, from 40 to 25 μm, and Γ, from 1 to 0.7. This gives $A\varphi$ = 240.53 μm² and $C\varphi = 9.70 \times 10^5$ μm² m N⁻¹. The next-highest point remains the Ls = 150 μm SiN string, at $C\varphi$ = 2540.45 μm² m N⁻¹. The trampoline value is higher by a factor of 381.77, or 382 after rounding.

The second test uses circular apertures constrained by the transverse width of each moving region. For the non-drum comparison devices, we set $\Gamma$=1 to obtain an upper bound on the phase-coupled area within each chosen aperture. The 2D-material drums retain their reported circular areas and $\Gamma$=0.432. In this test, the next-highest point is the square SiN membrane at $C_\phi$=378.863 μm² m N⁻¹. The trampoline value remains higher by a factor of $1.337\times10^4$, or $1.34\times10^4$ after rounding.

**Table S9. Display-rounded mechanical coefficients, phase-coupled areas, and coordinates for Fig. 2.**

| Point / reference | $K$ (N m⁻¹) | $K_3$ (N m⁻³) | $z_1$% (nm) | $A\varphi$ (μm²) | $C\varphi$ (μm² m N⁻¹) |
|---|---|---|---|---|---|
| Serpentine trampoline, M1 [4] | $2.48 \times 10^{-4}$ | $1.83 \times 10^{6}$ | 675.5 | 1257 | $5.067 \times 10^{6}$ |
| SiN nanowire pair, wide-wire mode [5] | 0.1 | $1.3 \times 10^{13}$ | 5.089 | 1.816 | 18.16 |
| Graphene drum, device 1 [6] | 0.5240 | $1.355 \times 10^{15}$ | 1.141 | 3.664 | 6.993 |
| $MoS_2$ drum, device 2 [6] | 1.077 | $1.252 \times 10^{15}$ | 1.702 | 2.345 | 2.177 |
| $MoS_2$ drum, device 3 [6] | 1.028 | $1.192 \times 10^{15}$ | 1.704 | 2.345 | 2.280 |
| Multilayer graphene drum [7,8] | 3.855 | $1.670 \times 10^{14}$ | 8.815 | 2.345 | 0.6083 |
| Single-layer graphene drum [9] | 0.01415 | $4.529 \times 10^{11}$ | 10.26 | 3.664 | 258.9 |
| SiN string, Ls = 30 μm [10] | 0.8522 | $1.745 \times 10^{10}$ | 405.5 | 162.1 | 190.2 |
| SiN string, Ls = 70 μm [10] | 0.2182 | $4.469 \times 10^{9}$ | 405.5 | 162.1 | 743 |
| SiN string, Ls = 110 μm [10] | 0.09881 | $2.024 \times 10^{9}$ | 405.5 | 162.1 | 1641 |
| SiN string, Ls = 150 μm [10] | 0.06381 | $1.307 \times 10^{9}$ | 405.5 | 162.1 | 2540 |
| Square SiN membrane, mode (1,1) [13] | 518.3 | $3.962 \times 10^{11}$ | 2099 | $4.106 \times 10^{4}$ | 79.23 |
| SiN beam, $L$ = 50 μm, n = 1 [11] | 9.05 | $5.315 \times 10^{12}$ | 75.72 | 18.24 | 2.015 |
| SiN beam, $L$ = 30 μm, n = 1 [11] | 2.41 | $1.596 \times 10^{13}$ | 22.55 | 10.94 | 4.541 |
| SiN beam A1 [12] | 2.2 | $4.065 \times 10^{13}$ | 13.50 | 1.52 | 0.6908 |
| SiN beam A2 [12] | 11 | $3.667 \times 10^{13}$ | 31.78 | 1.52 | 0.1382 |
| Graphene ribbon [15] | 0.3085 | $7.0 \times 10^{15}$ | 0.3852 | 0.08268 | 0.2680 |
| Carbon nanotube [15] | 0.01009 | $3.0 \times 10^{12}$ | 3.366 | $1.362 \times 10^{-3}$ | 0.1349 |
| Tapered Si nanowire, device 1 [14] | $1.57 \times 10^{-4}$ | $6.895 \times 10^{4}$ | 2769 | 0.333 | 2122 |

| **Point / reference** | ***K*** **(N m⁻¹)** | ***K₃*** **(N m⁻³)** | **z₁% (nm)** | ***Aφ*** **(μm²)** | ***Cφ*** **(μm² m N⁻¹)** |
|---|---|---|---|---|---|
| Tapered Si nanowire, device 2 [14] | $1.733 \times 10^{-4}$ | $7.38 \times 10^{4}$ | 2812 | 0.3812 | 2199 |
| Tapered Si nanowire, device 3 [14] | $2.814 \times 10^{-4}$ | $1.301 \times 10^{5}$ | 2699 | 0.558 | 1983 |
| Tapered Si nanowire, device 4 [14] | $5.971 \times 10^{-4}$ | $8.008 \times 10^{5}$ | 1584 | 0.328 | 549.3 |
| Tapered Si nanowire, device 5 [14] | $4.898 \times 10^{-3}$ | $1.354 \times 10^{7}$ | 1103 | 0.5359 | 109.4 |
| Tapered Si nanowire, device 6 [14] | $1.109 \times 10^{-3}$ | $1.511 \times 10^{6}$ | 1572 | 0.4242 | 382.4 |
| Tapered Si nanowire, device 7 [14] | $7.355 \times 10^{-4}$ | $1.05 \times 10^{6}$ | 1536 | 0.339 | 460.9 |
| Tapered Si nanowire, device 8 [14] | $3.882 \times 10^{-3}$ | $4.697 \times 10^{6}$ | 1668 | 0.8325 | 214.4 |
| Tapered Si nanowire, device 9 [14] | $6.228 \times 10^{-3}$ | $7.406 \times 10^{6}$ | 1683 | 1.04 | 166.9 |
| Tapered Si nanowire, device 10 [14] | $3.38 \times 10^{-3}$ | $9.515 \times 10^{6}$ | 1094 | 0.4486 | 132.7 |
| Tapered Si nanowire, device 11 [14] | 0.02032 | $5.264 \times 10^{7}$ | 1140 | 0.8767 | 43.15 |
| Tapered Si nanowire, device 12 [14] | 0.02553 | $5.809 \times 10^{7}$ | 1217 | 1.065 | 41.7 |
| Tapered Si nanowire, device 13 [14] | $4.185 \times 10^{-3}$ | $3.446 \times 10^{7}$ | 639.5 | 0.1757 | 41.99 |
| Tapered Si nanowire, device 14 [14] | 0.0154 | $1.234 \times 10^{8}$ | 648.3 | 0.3057 | 19.85 |
| Tapered Si nanowire, device 15 [14] | 0.1193 | $7.405 \times 10^{8}$ | 736.4 | 0.7097 | 5.951 |
| Tapered Si nanowire, device 16 [14] | 0.05107 | $1.133 \times 10^{8}$ | 1232 | 1.362 | 26.68 |
| Tapered Si nanowire, device 17 [14] | 0.01018 | $7.443 \times 10^{7}$ | 678.7 | 0.2968 | 29.15 |
| Tapered Si nanowire, device 18 [14] | 0.01698 | $1.251 \times 10^{8}$ | 675.9 | 0.3437 | 20.24 |
| Tapered Si nanowire, device 19 [14] | 0.01636 | $1.261 \times 10^{8}$ | 660.8 | 0.3295 | 20.14 |
| Tapered Si nanowire, device 20 [14] | 0.1189 | $8.221 \times 10^{8}$ | 697.8 | 0.6412 | 5.393 |

**S8. Co-design for accessing the predicted phase range**

The serpentine trampoline provides a quantitative starting point for designing a nonlinear optical phase element. This section uses the predicted $2\pi$ operating point to identify thermal design requirements and briefly describes the measured mechanical dynamics.

The mechanical parameters are from Ref. [4], while the optical and thermal inputs are from the closely related device in Ref. [3]. Both have a 50-nm-thick SiN pad with a serpentine suspension, but their spring geometries differ. The thermal analysis therefore gives reference-geometry scaling for a future design rather than an operating-temperature prediction for the Ref. [4] device.

**Table S10. Inputs for the thermal design analysis.**

| Symbol and quantity | Value | Source and role |
|---|---|---|
| Projected 2π power, $P_{2\pi}$ | 23.9 mW | Supplementary Section S4, Eq. (S11) |
| Reference absorptance, $A_{abs}$ | $4.1 \times 10^{-4}$ | Ref. [3], Supplementary Note 3, Eq. (S20) and following paragraph |
| Unfolded tether length, $L$ | 440 μm | Ref. [3], Supplementary Note 3, paragraph following Eq. (S20) |
| Tether cross section, $w \times t$ | 2 μm × 50 nm | Ref. [3], Supplementary Note 3, paragraph following Eq. (S20) |
| Thermal conductivity, $\kappa_{th}$ | 2.5 W m$^{-1}$ $K^{-1}$ | Ref. [3], Supplementary Note 3, paragraph following Eq. (S20) |

For incident power $P$ and absorptance $A_{\text{abs}}$, the absorbed power is

$$P_{\text{abs}} = A_{\text{abs}} P. \tag{S31}$$

For four identical tethers, the conduction model of Ref. [3], Supplementary Note 3, Eqs. (S13)–(S14), gives

$$G_{\text{th}} = \frac{4\kappa_{\text{th}} w t}{L}. \tag{S32}$$

Here $\kappa_{\text{th}}$, $w$, $t$, and $L$ are the thermal conductivity, width, thickness, and unfolded length of each tether. The inputs in Table S10 give $P_{\text{abs}}$=9.8 μW at $P_{2\pi}$=23.9 mW, and a total thermal conductance $G_{\text{th}}$=2.27 nW K$^{-1}$.

Within the small-temperature regime of the constant-property conduction model, $\Delta T = A_{\text{abs}} P / G_{\text{th}}$. For a chosen allowable temperature rise $\Delta T_{\max}$ relative to the substrate, accessing the full phase range therefore requires

$$\frac{G_{\text{th}}}{A_{\text{abs}}} \geq \frac{P_{2\pi}}{\Delta T_{\max}}. \tag{S33}$$

For an illustrative $\Delta T_{\max}$=10 K, the required ratio is $G_{\text{th}}/A_{\text{abs}} \geq 2.39\times10^{-3}$ W K$^{-1}$, about 430 times the reference value of $5.54\times10^{-6}$ W K$^{-1}$. This is a joint target involving absorptance, thermal conductance, and the incident power required for the phase shift.

The required incident power depends on the optical momentum transfer and mechanical restoring force. Using the notation of Section S1, a full $2\pi$ reflected-phase excursion requires

$$\begin{aligned} P_{2\pi} &= \frac{c}{\xi \Gamma_d} F_m\left(\frac{\lambda}{2\Gamma_r}\right), \\ P_{2\pi} &\approx \frac{cK\lambda}{2\xi \Gamma_d \Gamma_r}, \quad F_m(z) \approx Kz. \end{aligned} \tag{S34}$$

Here $F_m(z)$ is the restoring force, $c$ is the speed of light, $\lambda$ is the wavelength, and $\Gamma_d$ and $\Gamma_r$ are the drive and readout overlaps. The approximation uses the linear stiffness $K$.

Lower absorptance and greater optical momentum transfer reduce the thermal burden. Changes to the suspension can improve conduction but also change the linear and cubic restoring-force coefficients. The coupled design objective is therefore to reduce $A_{\mathrm{abs}}P_{2\pi}/G_{\mathrm{th}}$ while retaining the desired phase responsivity and range.

Land et al. [19] inferred extinction coefficients of approximately 0.1–1 ppm for stoichiometric SiN at wavelengths from 532 to 1550 nm. These coefficients are distinct from device absorptance. Converting them into $A_{\mathrm{abs}}$ requires an optical calculation for the chosen membrane thickness and structure.

The source trampoline has $f_0$=5,862 Hz and $Q$=49,420 [4], characterizing its weakly damped dynamics. The temporal behavior desired of a phase element depends on its application, including the balance between response to changing illumination and persistence of mechanical motion. Evaluating that behavior requires the damping and thermal properties of a chosen design.

**Supplementary references**